# Advanced piezoresistance of extended metal/insulator core shell nanoparticle assemblies

*E. K. Athanassiou [1], F. Krumeich [2], R. N. Grass [1] and W. J. Stark [*1]*

[1] Institute for Chemical and Bioengineering, ETH Zürich, Zurich, CH-8093 Switzerland

[2] Laboratory of Inorganic Chemistry, ETH Zürich, Zurich, CH-8093 Switzerland

[*] Correspondence should be addressed to W. J. Stark

Email: wendelin.stark@chem.ethz.ch

Fax: +41 44 633 10 83

**Abstract**

Assembled metal/insulator nanoparticles with a core/shell geometry provide access to materials containing a large number ($>10^6$) of tunneling barriers. We demonstrate the production of ceramic coated metal nanoparticles exhibiting an exceptional pressure sensitive conductivity. We further show that graphene bi- and trilayers on 20 nm copper nanoparticles are insulating in such core/shell geometry and show a similar pressure dependent conductivity. This demonstrates that core/shell metal/insulator assemblies offer a route to alternative sensing materials.

Classical electron tunneling through an insulating gap is highly sensitive both to barrier thickness and height. Since real materials have a non-negligible compressibility, such an arrangement can be viewed as the most primitive realization of a pressure sensor. The demanded technical accuracy of such systems [1-5], however, would be far from economically interesting as changes in the picometer range already strongly affect the tunneling properties of a typical single barrier. A potential solution around this yet unsolved manufacturing problem comes from the use of large, serial arrays of tunneling barriers where an overall material property would be the combined result of literally millions of tunneling processes. Measuring a distribution of tunneling properties then provides a technically feasible route to use the extreme sensitivity of a barrier for pressure sensing.

We have most recently reported on the unexpected high piezoresistance and thermoresistive effect of graphene bi- and trilayer coated copper nanoparticles [3] and suggested possible explanations based on multielectron tunneling mechanism derived from the temperature dependence of the electrical resistivity [6]. In contrast, multilayer graphene coated (4-6 layers) copper nanoparticles exhibited a much higher conductivity but much weaker thermo- or piezoresistance effects. The puzzling insulating properties of the graphene bi- and tri-layers on the 20 to 50 nm sized copper nanoparticles [6, 7] escaped any attempts for explanation until the most recent appearance of a study by Oostinga et al. who confirmed that graphene bi-layers could become insulating upon application of a strong electrical field perpendicular to the sheet's orientation [5] or due to adsorption of metallic potassium [8]. Combining these two most recent studies provides the motivation for the present work where we experimentally proofed that core/shell metal/insulator nanoparticle arrays offer a route to highly sensitive, non-classical piezo- and thermoresistive materials.

Ceramic coated metal nanoparticles would provide a most natural choice for a core/shell metal/insulator material but the different surface energy of most metal/ceramic pairs generally prohibits the formation of ceramic shells around metal nanoparticles. This was most recently shown by Grass et al. [9] who simultaneously prepared metal and ceramic nanoparticles in a high temperature aerosol process but observed formation of separate particles of metal and ceramic. Combining a more advanced ceramic (zirconia-doped ceria) with a suitable metal alloy (Ni/Mo, superalloy) in the present work, however, gives access to a material pair with stable ceramic coatings on metal cores. The wetting is facilitated through the high solubility for nickel and molybdenum [14] in the here chosen ceramic. Such oxides find broad application in heterogeneous catalysts, particularly where dynamic reduction is important, e.g. in car exhaust gas treatment [11, 12]. Our observations are in line with Sun et al. [10] who have previously reported on the encapsulation of palladium by ceria/zirconia.

Nickel molybdenum alloy nanoparticles (Ni/Mo = 7/3 at/at) coated with ceria/zirconia (Ce:Zr ratio = 1:1 at/at; 10 vol % overall) were produced in a single step at a production rate of about 20 g h$^{-1}$ by reducing flame synthesis [13] (see EPAPS Document No. for the production and detailed physical analysis of NiMo/Ceramic). Transmission electron micrographs displayed nearly spherical particles with a surface coating [Fig. 1(a)] which stays in agreement with the air stability and the previous synthesis of bilayer carbon coated copper nanoparticles [Fig. 1(b); 2 C$_\infty$/Cu]. The preferential reduction of nickel and molybdenum to their metallic form and the simultaneous formation of cerium and zirconium oxide are in agreement with the large difference in the reduction potential of the composite constituents. The ceramic coating has been successfully formed round the nickel/molybdenum nanoparticles and did not have any measurable influence on the physical properties of the metal core as thermogravimetric analysis has showed (see EPAPS Document No.; [Fig. S5]). The here observed improved wetting of the ceramic can be traced to partial solubility of nickel and molybdenum oxide inside ceria/zirconia ceramics [14]. The black core/shell metal/insulator powder was pressed to cubic objects (10 mm x 10 mm x 1.2 mm) by uniaxial compression and the piezoresistivity measurements of the NiMo/Ceramic assemblies [Fig. 1(c)] were performed under uniaxial dynamic pressure as described in the previous study [6]. Full reversibility of the piezo-resistive effect was confirmed by applying pressure cycles (see EPAPS Document No.; [Fig. S6]). The electrical resistivity rapidly decreased upon increasing pressure and showed a strong power law correlation over three orders of magnitude [Fig. 2, NiMo/Ceramic; $x^{-0.71}$]. This technically attractive value compares favorably to currently used conductive polymer systems or carbon nanotubes [4, 15, 16] and a very different core/shell material we most recently prepared in our laboratory. Pressed assemblies of graphene bi- and trilayer coated 20-50 nm size copper nanoparticles [Fig. 1(a)] [3] most recently showed an astonishingly similar pressure dependent conductivity (Fig. 2; 2 $C_\infty$/Cu; $x^{-0.78}$). Even though the absolute conductivity of the two nanocomposites was very different, they exhibit a most similar sensitive negative pressure-coefficient effect of resistance (NPCR [16-18]) in contrast to multilayer graphene coated particles from previous studies (4-6 $C_\infty$/Cu; $x^{-0.54}$, [6]).

We can compare the here-prepared material to engineered polymer composites containing fibrous fillers such as graphite sheets or carbon black where polymer deformation can result in a (reversible) breakdown of the conductive filler network and a positive pressure coefficient effect of resistivity (PPCR [17, 18]). In contrast, conduction in polymers with metal fillers is based on a pressure dependent reduction of the metal particle/particle distance and a resulting increase in conductive path density with a negative pressure coefficient effect of resistance (NPCR) [17, 18]. This effect is qualitatively similar to the here-observed core/shell metal/insulator assemblies [3, 6] and numerous conductive filler materials have been reported with a power law correlation [4, 15, 19, 20] over a wide range of pressure ranges. The use of 0.6 to about 1 nm thick graphene bi- and trilayers can be considered as the ultimate design of a sandwich construction based on a metal conductor framework with interdispersed, compressible poor- or non-conducting barriers.

The conduction mechanism between adjacent particles arises from tunneling processes [21-24] and can be described as:

$$\sigma_c = \sigma_f \exp(-2X_t d) \quad (1)$$

where $\sigma_c$ and $\sigma_f$ are the conductivities of the composite and the conductive fillers, respectively, d is the distance between the particles and $X_t$ is defined as $X_t = [8\pi^2 m V(t)/h^2]^{0.5}$ with the mass of the charge carriers $m$ and $V(t)$ as the temperature modified barrier height. Tunneling takes place exclusively between neighboring particles. Even if the particles are in direct physical contact [25], conductive chains (Scheme 1, right) contain gaps across which electrons can tunnel. The pressure dependence comes from the compressibility of the materials which affects the barrier width $d$ (Scheme 1, left, with d as the interparticle or particle/particle distance) and can be described by $d=d_0(1-\varepsilon)= d_0(1-P/M)$, where $d_o$ is the interparticle gap width in the original sample, $\varepsilon$ the strain of the sample, $P$ the applied uniaxial pressure and $M$ the bulk compressive modulus of the shell. Under uniaxial pressure, we can combine these expressions to:

$$\rho(P) = \rho_o \exp(-2X_t d_o P/M) \quad (2)$$

where $\rho(P)$ is the composite resistivity under pressure and $\rho_o$ is the resistivity of the original composite sample [6]. Concerning that only under high applied pressures the neighboring particles have a physical contact we compared the corresponding experimental data of NiMo/Ceramic to the tunneling model (see EPAPS Document No.; [Fig. S7]). Comparison to our most recently reported highly pressure dependent conductivity of graphene coated copper nanoparticles [3] further experimentally confirms the similarity between NiMo/Ceramic and 2 $C_\infty$/Cu (Table 1). Assuming that the compressive modulus of stacked graphene layers (graphite) perpendicular to the planes is $M$ = 2 GPa [26] and for ceria $M$ = 4.2 GPa [27] and that the interparticle distance is twice the thickness of the graphene or ceramic coating (Table 1) for neighboring particles, the term $mV(t)$ can be calculated. Although there are many studies determining the mass of charge carriers along a graphene plane, no information was found about the mass of charge carriers perpendicular to a graphene or ceramic plane. Assuming that $m= 0.5\ m_e$ an energy barrier of $V(t)$ = 1.46 eV for NiMo/Ceramic and $V(t)$ = 1.68 eV for 2 $C_\infty$/Cu has been calculated (Table 1). This observation further confirms the pronounced dependence of the piezoresistance on the architecture of such a multi-barrier (core/shell) system and the shell's electronic properties. In contrast to these samples, the calculated barrier height of 4-6 $C_\infty$/Cu was much lower ($V(t)$ = 0.09 eV, Table 1) explaining the almost absent sensitivity of 4-6 $C_\infty$/Cu and reflecting the conductive behavior of the graphene stacks (4-6 layers) covering the copper nanoparticles.

The low conductivity of the shells in both ceramic/metal and bilayer carbon/copper is not astonishing in case of the ceria/zirconia oxide layer but graphene-type carbon is a semimetal (zero-bandgap semiconductor) with considerable electrical conductivity. Hence, one would not expect it to serve as a compressible insulating surface layer [Fig. 1(b)]. The most recent study by Oostinga et al. [5] shows that graphene bilayers can become insulating upon application of strong electric fields perpendicular to the graphene sheets by induction of an energy gap between the conduction and valence band. Latter has been proved through simulations studies for a bilayer graphene by Mc Cann [28], for graphene nanoribbons [29, 30] or for conductor/graphene/conductor junctions [31]. The presence of defects or charge impurities on the layer surface could also influence the conductivity of the graphene shell [32-34]. Theoretical and experimental studies by Kim et al. have showed that the epitaxial growth of graphene on a silicon carbide surface induces a gap opening in midgap states at the Dirac point of graphene. The gap originates from sublattice symmetry breaking interactions between graphene and the interfacial structure in the system [35] indicating that the here investigated core/shell geometry results in a similar effect. The carrier density in graphene bilayers and the occupation of electronic states near the Fermi level could be experimentally manipulated within a study by Ohta et al. [8] through the surface adsorption of metallic potassium. That study suggests that the here discussed graphene bilayers on 20 nm copper nanoparticles [Fig. 1(b)] could behave similarly. This would account for the very low conductivity of the graphene due to the presence of an energy barrier (Scheme 1, left) in these samples and give a reasonable explanation to why core/shell metal/insulator materials can be easily prepared from the assembly of either ceramic coated metal or graphene bilayer coated copper nanoparticles. A series of studies [36-38] have shown that bilayer and multilayer (more than three layers) graphene exhibit different electronic properties. While graphene bilayers are zero-gap semiconductors, the conduction and valence band of multilayer graphene strongly overlap making it similar to a thin film of graphite. We can now further re-interpret our most recent investigation [6] comparing the pressure dependent conductivity of 20 nm copper nanoparticles enveloped in graphene bilayers (excellent sensing properties, $d_{Layer\ thickness}$ = 0.8 nm, Table 1) to graphene multilayers with 4-6 carbon sheets (no sensing properties observed [6]). The thicker graphene layer ($d_{Layer\ thickness}$ = 1.8 nm; 4-6 layers, Table 1) could not be insulating [5, 8] because the top graphene layers remain in a classical, conducting graphene state. The present study showed how a pronounced piezoresistance effect could be realized using an assembly of tunneling barriers. The tunneling barrier sensitivity could be engineered by preparing two different core/shell materials exhibiting different electrical conductivity but similar piezoresistance. The conductivity strongly depends on the geometry and the physical properties of the shell layer. The pressure dependence of here described material's electrical properties is most presumably based on tunneling processes and may be subject to the Klein paradox [39], but this aspect requires further investigation.

Large scale assemblies of core/shell metal/insulator nanoparticles have been demonstrated to offer an experimentally easily accessible way to materials with high pressure sensing properties (Scheme 1, right). The assembly links a local property (tunneling) to a macroscopic output (electrical conductivity). Our study therefore showed how a macroscopic property enables the measurement of a large distribution of parallel tunneling processes in a material and suggests the use of such assemblies for new sensing methods.

The authors thank the Electron Microscopy Center of ETH Zurich (EMEZ) for the transmission microscopy imaging and energy dispersive X-ray spectroscopy. Financial support by the ETH Zurich (TH 02-07-3) and the Swiss National Science Foundation (SNF200021-116123) are gratefully acknowledged.

**TABLES**

**TABLE 1.** Particle size, layer thickness and barrier height of NiMo/Ceramic and C/Cu samples.

| Samples | $d_{Metal\ core}$ (nm)[a] | $d_{Layer\ thickness}$ (nm)[b] | Layer properties | $V(t)$(eV) | Ref. |
|---|---|---|---|---|---|
| NiMo/Ceramic | 16.4 | 1.2 | Ceramic insulating | 1.46[c] | This work |
| 2 $C_\infty$/Cu | 10.3 | 0.8 | Graphene bilayer | 1.68[d] | [6] |
| 4-6 $C_\infty$/Cu | 12.7 | 1.7 | Graphene stack | 0.09[d] | [6] |

[a] Particle size calculated from the specific surface area, error ± 10% (Assumed $\rho_{Ni/Mo}$: 8.9 g cm$^{-3}$, $\rho_{Cu}$: 8.9 g cm$^{-3}$, [b] Layer thickness determined by TEM images, error ± 15%, [c] Assuming $M_{ceramic}$ = 4.2 GPa and $m$ = 0.5 $m_e$, [d] Assuming $M_{graphene}$ = 2.0 GPa and $m$ = 0.5 $m_e$

**Figure Captions**

FIG. 1. (a) Transmission electron micrographs (TEM) of NiMo/ceramic showing almost spherical 10-50 nm nanoparticles covered by an amorphous oxide layer. (b) Copper nanoparticles covered by thin graphene layers. (c) Macroscopic assembly of core shell NiMo/Ceramic nanoparticles.

FIG. 2. The pressure dependence of the electrical conductivity of NiMo/ceramic showed a similar piezoelectric behavior as the insulating carbon coated copper ($2C_\infty$/Cu) nanomaterials. In spite of the better conductivity of NiMo/Ceramic the resistivity of both materials followed a power law dependence (NiMo/Ceramic: $x^{-0.71}$; $2C_\infty$/Cu: $x^{-0.78}$) confirming a high piezoresistance effect in such core/shell assemblies.

Scheme 1. (left) Two core/shell metal/insulator nanoparticles in close vicinity can be viewed as a classical tunneling barrier situation. The energy gap arises from the different Fermi levels in the metal cores and the shell material (ceramic or a graphene bi- or trilayers). (right) Electron transport through an assembly of such core shell composites follows multiple parallel path along particle chains. The macroscopic conductivity is a result of a large number of parallel and serial tunneling gaps along the conduction path.

**Figures**

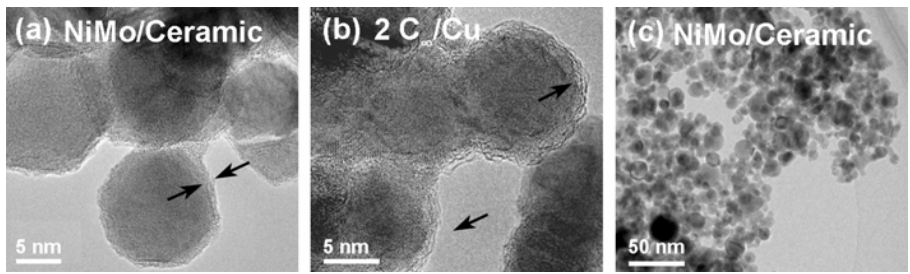

FIG. 1.

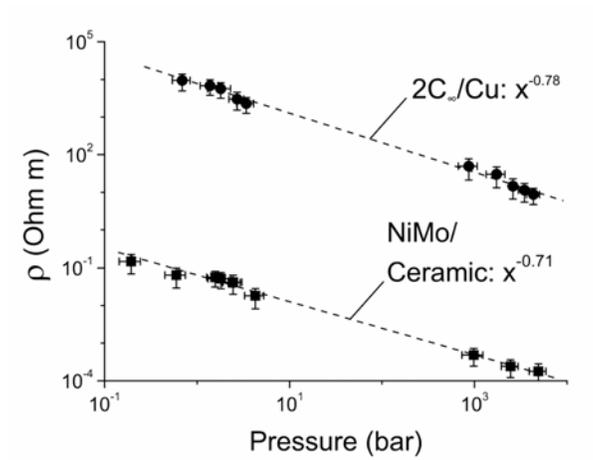

FIG. 2.

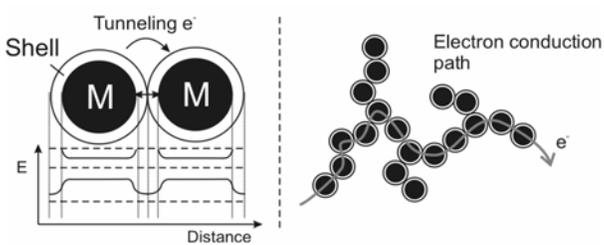

Scheme 1.